\documentclass [epsfig,12pt,floatfix,amsmath,amssym]{revtex4}

\usepackage[pdftex]{graphicx}
\usepackage{color} 
\begin{document}

\newcommand{\xv}{{\mathbf x}}
\newcommand{\rv}{{\bf r}}
\newcommand{\uv}{{\bf u}}
\newcommand{\tv}{\hat{{\bf t}}}
\newcommand{\nv}{{\bf n}}
\newcommand{\vv}{{\bf v}}
\newcommand{\kv}{{\bf k}}
\newcommand{\mv}{{\bf m}}
\newcommand{\Uv}{{\bf U}}
\newcommand{\Vv}{{\bf V}}
\newcommand{\Lm}{{\bf \Lambda}}
\newcommand{\Mm}{{\bf M}}
\newcommand{\Pm}{{\bf P}}
\newcommand{\qv}{{\mathbf q}}
\newcommand{\av}{{\mathbf a}}
\newcommand{\cH}{{\cal H}}
\newcommand{\cZ}{{\cal Z}}
\newcommand{\zh}{\hat{z}}
\newcommand{\ph}{\hat{\phi}}
\newcommand{\rh}{\hat{r}}
\newcommand{\ev}{\hat{{\bf e}}}
\newcommand{\lB}{\ell_B}
\newcommand{\grad}{{\bf \nabla}}
\newcommand{\bi}{\bibitem}

\title{Conformational Collapse of Surface-Bound Helical Filaments}

\author{David A. Quint$^{1}$,  Ajay Gopinathan$^{1 \ast}$ and Gregory M. Grason$^{2\dagger}$}

\address{$^1$Department of Physics, University of California, Merced, CA 95343 USA}
\address{$^2$Department of Polymer Science and Engineering, University of Massachusetts, Amherst, Massachusetts 01003, USA}
\address{ E-mail:  $^\ast$ agopinathan@ucmerced.edu; $^\dagger$ grason@mail.pse.umass.edu}

\begin{abstract}
Chiral polymers are ubiquitous in nature and in the cellular context they are often found in association with membranes. Here we show that surface bound polymers with an intrinsic twist and anisotropic bending stiffness can exhibit a sharp continuous phase transition between states with very different effective persistence lengths as the binding affinity is increased. Above a critical value for the binding strength, determined solely by the torsional modulus and intrinsic twist rate, the filament can exist in a zero twist, surface bound state with a homogeneous stiffness. Below the critical binding strength, twist walls proliferate and function as weak or floppy joints that sharply reduce the effective persistence length that is measurable on long lengthscales. The existence of such dramatically different conformational states has implications for both biopolymer function {\it in vivo} and for experimental observations of such filaments {\it in vitro}.
\end{abstract}

\maketitle

\section{Introduction}
\label{sec:intro}
Both prokaryotic and eukaryotic cells possess a rich variety of  filamentous proteins with a wide range of compositions, sizes and flexibility \cite{lodish}.  These biopolymers play a vital role in a number of critical cellular functions including maintaining structural integrity, serving as a template for cell growth, cell adhesion and motility, mechanical signal transduction and cell division. In many of these roles, the association of the polymers with membranes is critical and necessary for function. For example, the actin rich cell cortex in eukaryotic cells is a major determinant of the cell's structural integrity and mediates adhesion and signal transduction. A number of different actin binding proteins link the cortex with the membrane both directly, and indirectly through other proteins and complexes \cite{cowin}. In the axons and dendrites of nerve cells, members of the MAP1 family link microtubules, which are critical for vesicle transport and mechanotransduction, to the plasma membrane \cite{mts}. In red blood cells, the remarkably flexible spectrin-actin network, responsible for cell shape regulation, is tightly associated with the membrane by a number of proteins such as ankyrin and dematin \cite{janmey}. Similarly, the structural support lamin networks inside the nucleus are attached to the nuclear envelope by LAP family proteins \cite{lamin}. During cell division, forces required for pinching in the division plane are generated by shrinking actin contractile rings attached to the membrane \cite{cont}. A similar phenomena occurs in bacterial cell division where the bacterial filamentous protein FtsZ makes up the contractile ring \cite{ftsz} which can be linked to the membrane by FtsA, ZipA and ZapA \cite{ftsa,arkin}. Another bacterial protein MreB which provides structural support and guides cell wall synthesis is attached to the membrane via RodZ, MreC/D \cite{mreb,arkin} or even directly \cite{mrebdir}. While these and several more examples illustrate the common nature of filamentous proteins binding to membranes {\it in vivo}, it is also known that many filamentous proteins can show direct interactions with membrane components {\it in vitro} typically through electrostatic or hydrophobic interactions \cite{janmey}. This is of fundamental importance when considers that many experimental setups that study such filaments {\it in vitro} have them in close proximity or immobilized/fixed on a surface. Clearly, understanding the association of filamentous proteins with membranes and surfaces is important from a scientific and technological perspective. \par
To date, there has a been a large amount of work studying polymers at interfaces and surfaces \cite{polyint,polyint2}. However, many of these studies tend to neglect key features of real biopolymers that could have qualitative effects on their structure and their interactions with surfaces. Firstly, a vast majority of biopolymers are chiral with an intrinsically helical structure. This helicity can manifest itself in a number of fascinating ways such as the cholesteric liquid crystal and the blue phases of concentrated DNA solutions and the self-limiting bundle size of aggregates of generic chiral biopolymers such as actin \cite{greg1,greg2,greg3,greg4}. Furthermore, classical models of polymers such as the worm-like chain model do not take anisotropies in elastic moduli or finite widths of the biopolymers into account. A number of recent efforts have therefore focused on the statistical mechanics of elastic ``ribbons'' with a finite aspect ratio as a more realistic model for typical biopolymers or other protein aggregates \cite{1rib1,1rib2,1rib3,1rib4,1rib5,1rib6,1rib7,1rib8,1rib9,1rib10,1rib11}. The ribbon-like nature of the polymers can lead to interesting effects including layering transitions in highly anisotropic condensates and the existence of an underlying helical structure even in the absence of twist \cite{2rib1,2rib2}. Thus the introduction of helicity and anisotropy can result in qualitative changes in the structure and dynamics of both single filaments and their aggregates.\par

Given the ubiquity of membrane and surface adsorbed biopolymers, in this paper, we take the first step toward understanding how helicity and anisotropy can influence the structure and dynamics of polymers that are bound to surfaces. Rather than study a polymer with more detailed resolution, for example a full atomistic representation, we focus on the generic effects  of intrinsic helicity on polymer adsorption using a minimal, coarse-grained description of molecular structure. We show that the interplay between helicity and surface interactions  can have profound effects on the structural and thermodynamic properties of the polymer.  We thus probe the consequences of helical structure on the large-length scale properties of absorbed chains, which is of potential relevance to a broad class of biomacromolecules. We first build a general model that allows us to prescribe the thermodynamics of a helical and anisotropic filament binding to a surface of arbitrary curvature. For the rest of the paper, we then treat only planar surfaces. We first consider the infinitely stiff limit where the anisotropy becomes irrelevant and only the helicity plays a role. 

The key conclusion from this analysis, is that the structure and conformational fluctuations of a helical polymer are critically sensitive to the attractive interactions with a surface.  We find that the filament undergoes a rapid and continuous phase transition from a strongly bound untwisted state to a weakly bound state for a critical value of the binding strength.The transition occurs via a proliferation of ``twist-walls'', regions where the chain unbinds locally from the surface and rapidly twists a single turn. We then consider how this transition alters the conformations and fluctuations of an anisotropic flexible filament on a planar surface. The coupling between the twist and bend degrees of freedom introduced by the anisotropy and binding results in surface bound conformations with highly heterogeneous bending stiffness below the critical binding strength. The transition between strongly and weakly bound states is then accompanied by a sharp drop in the effective bending rigidity of the polymer.  We demonstrate these conclusions, first, through an exact, ``zero-temperature" analysis of the continuum mechanical model of the filament that ignores thermal fluctuations of twist.  We then extend our analysis to finite-temperature by way of Monte Carlo simulations, which demonstrate that the sharp transition between the strongly-bound and weakly-bound states of helical filaments persists in the presence of both bending and torsion fluctuations of the chain backbone. Thus, we predict that strong coupling of the twist degrees of freedom of a helical polymer to surface interactions, not only restructures the torsional state of the chain, but it also leads to a dramatic transition in the conformational fluctuations of the entire chain backbone.  This suggests that experimentally measured properties of biopolymers absorbed to a surface such as the persistence length and end-to-end size are crucially sensitive to the strength of surface interactions.

\par
This paper is organized as follows. In Section.\ref{sec:model}, we build the general model for a helical and anisotropic filament binding to a surface of arbitrary curvature and in Section.\ref{sec:sims}, we describe our Monte Carlo simulation procedure. In Section.\ref{sec:twist}, we analyze the model for planar surfaces in the infinitely stiff filament limit. We specifically study the transition between the strongly bound and weakly bound states of the filament. In Section.\ref{sec:bend}, we study how the transition influences the in-plane bending fluctuations of an anisotropic surface adsorbed filament. We conclude with a short discussion in Section.\ref{sec:conclusion}.

\section{Continuum theory of surface-bound helical filament}

\subsection{Model}
\label{sec:model}
Here we introduce to a continuum model of a helical filament bound to an attractive surface.  For the purposes of generality, we consider the filament bound to a cylindrical surface of radius, $r$.  This is a natural geometry to describe the binding the cylindrical cell wall of a bacterium, as in \cite{arkin} , though we will specialize in the next section to the case of planar surfaces, $r \to \infty$.  

The filament is modelled by a helical strip, whose cross-section is rectangular with a thickness, $t$ and width, $w$.  In the absence of external forces or interactions with the surface, the wide axis of the filament has a natural rotation rate, $\omega_0$, along the long axis of the filament (see Fig.~\ref{fig: figure1}).  Both thermal energy and surface interactions distort the filament geometry, leading to elastic costs which we described with the Kirchoff-Love of elastic beams~\cite{landau}.  The deformations of the filament are described by rotations of the material frame along the arc-length, $s$, of the filament backbone. We choose $\ev_1$ ($\ev_2$) to describe the local orientation of the wide (thin) axis of the filament, and $\ev_3 = \tv$ to describe the tangent.  

\begin{figure}
\center \includegraphics[clip=true, trim=50 100 50 30,scale=0.4]{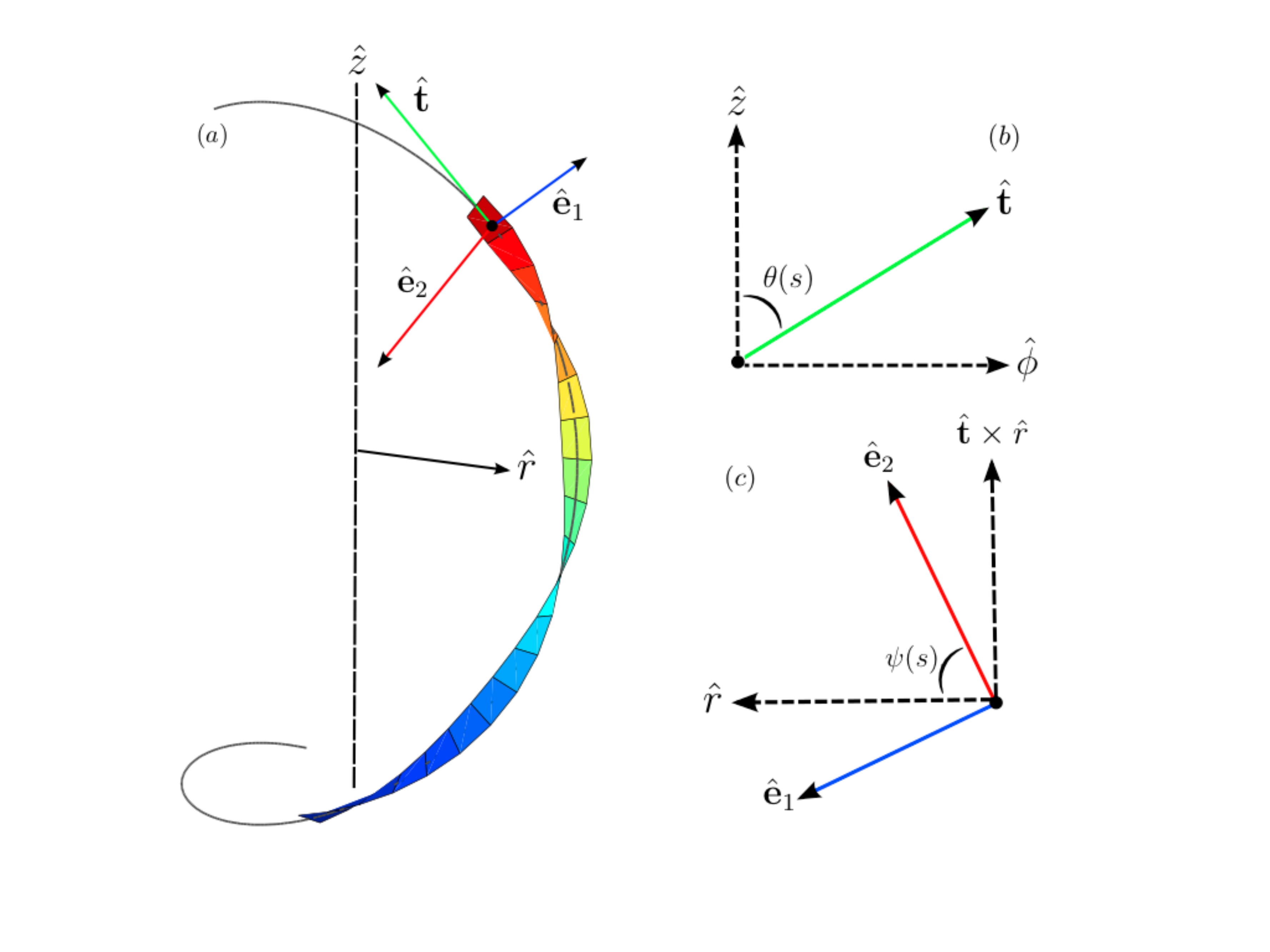} 
\caption{Diagram defining the coordinates used to describe an intrinsically helical filament. (a) A section of an intrinsically  twisted filament which is forming a helix in a cylindrical coordinate system, where $\tv$, $\ev_1$ and $\ev_2$ are the material frame basis vectors. The grey curve is the center line of the filament.  (b) The projection of the tangent vector, $\tv$, along $\hat{z}$ and $\hat{\phi}$ defines of the bending angle $\theta(s)$. (c) The rotation of the material coordinates $\ev_1$ and $\ev_2$ defines of the twist angle $\psi(s)$.}
\label{fig: figure0}
\end{figure}

\begin{figure}
\center \includegraphics[clip=true, trim=50 100 50 100,scale=0.4]{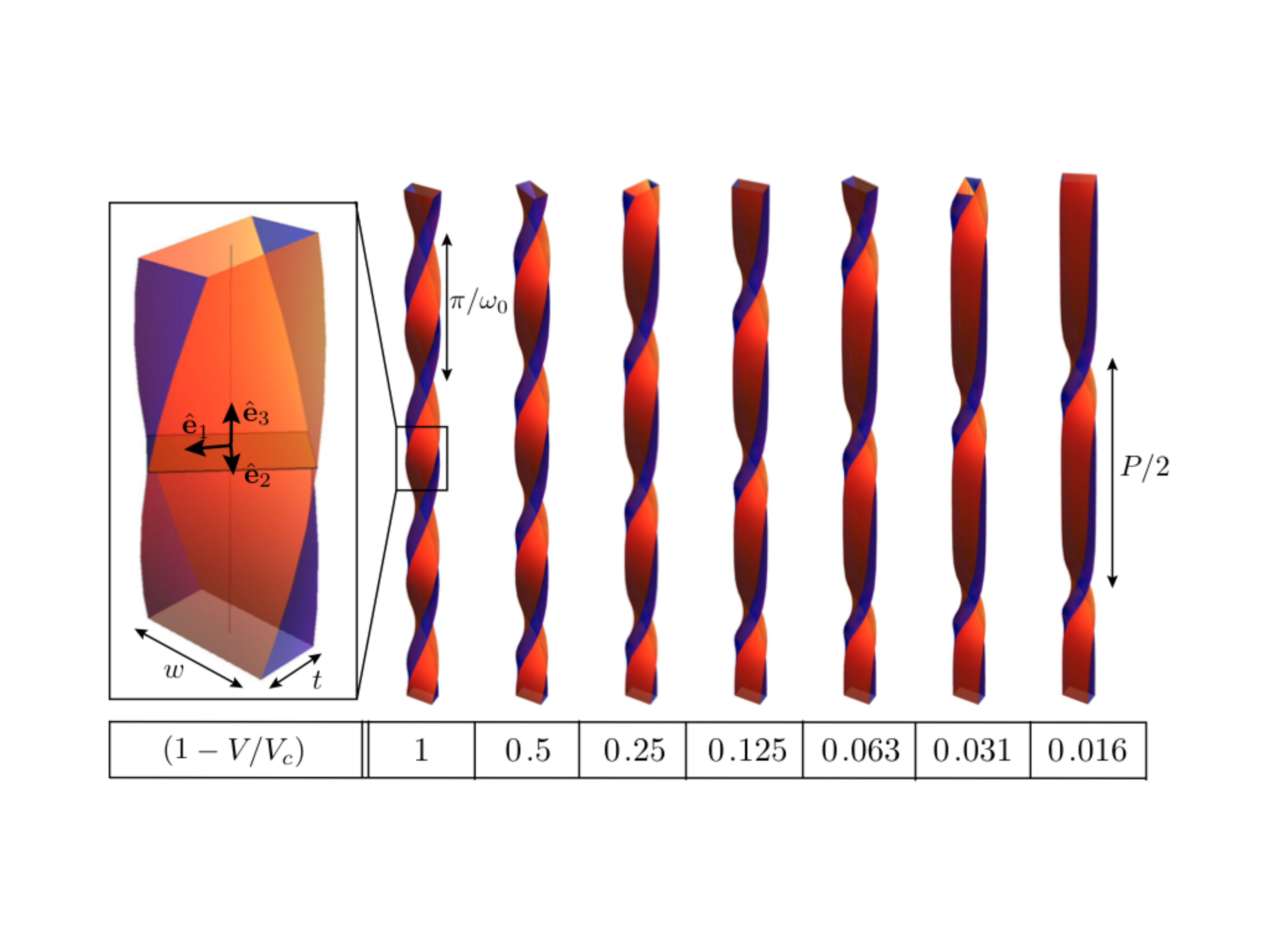} \caption{Schematic of a helical filament with an anisotropic cross section.  On the right, we show a series of helical configurations for increasing strength of surface interactions, which show progressive unwinding of the helical twist. }
\label{fig: figure1}
\end{figure}

Here, we compute the mechanical energy of the filament in terms of an angle, $\theta(s)$, describing in-plane orientation of $\tv$ and an angle, $\psi(s)$, describing the orientation of the wide axis of the filament with respect to the surface normal.  If the center line of the filament, ${\bf r}(s)$, is confined to a helical surface (with long axis centered on the $x=y=0$ axis), then the tangent vector can be described in terms of the angle $\theta(s)$, where
\begin{equation}
\partial_s {\bf r} = \tv = \cos \theta(s) \zh+ \sin \theta(s) \ph ,
\end{equation}
where $ \ph = \zh \times {\bf r}(s)/r$ is the local, azimuthal direction on the cylinder.  The curvature of the filament center is determined by $\partial_s \tv$, which is non-zero when $\theta' \neq0$ and the filament curves in the surfaces of the cylinder, or when $\sin \theta \neq0$, and the filament, following a straight, or geodesdic, path on the cylinder is deflected normal to the cylinder.  That is, the $\ph$ direction changes along the arc-length as the curve is convected around the cylinder axis as $
\partial_s \ph  =-\frac{\sin \theta}{r} \rh$ where the radial direction is given by $ \rh = [\rv - \zh (\rv \cdot \zh)]/r$ .  Hence,
\begin{equation}
\partial_s \tv = - \frac{ \sin^2 \theta}{r} \rh + \theta' (\tv \times \rh ),
\end{equation}
where $\tv \times \rh = \sin \theta \zh - \cos \theta \ph$.  To describe the local anisotropy of the filament cross-section we define the material directions, perpendicular to $\tv$,
\begin{equation}
\ev_1 = \sin \psi \rh - \cos \psi (\tv \times \rh )  ,
\end{equation}
and 
\begin{equation}
\ev_2 = \cos \psi \rh + \sin \psi (\tv \times \rh ) .
\end{equation}
Defining $\kappa_i = \ev_i \cdot \partial_s \tv$ as the rate of back bone bending into the direction $\ev_i$, we have,
\begin{equation}
\kappa_1 = \kappa \sin(\psi -\delta),  \ \kappa_2 = \kappa \cos(\psi - \delta),
\end{equation}
where 
\begin{equation}
\kappa^2 = (\theta ')^2 + \frac{ \sin^4 \theta}{r^2},
\end{equation}
is the square curvature of the central axis and $\tan \delta = r \theta'/\sin^2 \theta$.  Finally, we extract the rate of cross section twist, $\omega$, from the rate of rotation of $\ev_1$ and $\ev_2$ around the tangent,
\begin{equation}
\omega = \ev_2 \cdot \partial_s \ev_1= \psi' - \frac{\sin (2 \theta) }{2 r}.
\end{equation}
From these quantities we define the following elastic energy for the helical filament,
\begin{equation}
\label{eq: Emech}
E_{\rm mech} = \frac{1}{2} \int ds \Big[ C_1 \kappa_1^2 + C_2 \kappa_2^2 + K (\omega - \omega_0)^2 \Big] ,
\end{equation}
where $C_1$ and $C_2$ are bend moduli for bending around the respective wide and thin axes of the filament cross section and $K$ is the twist modulus.  Modeling the filament interior as an isotropic, elastic medium these three constants are related by
\begin{equation}
K =  \frac{ 3\mu (w t)^3}{\pi^2( w^2+t^2)}  ; C_1 = \frac{  E w^3 t}{12} ; C_2 = \frac{E w t^3}{12} 
\end{equation}
where $E$ and $\mu$ are the Young's and shear moduli of the of the elastic medium~\footnote{We have used approximation of the torsional modulus of a rectangular beam, $K = \mu A^4/ 4 \pi^2 I$, where $A$ and $I$ and the area and second-moment of the cross section, respectively.  See e.g. A. E. Love {\it A Treatise on the Mathematical Theory of Elasticity} (New York, Dover, 1944), 4th ed., Chap. 14.}.

Next we introduce an effective potential to describe the thermodynamics of interactions between the helical filament and the attractive wall.   We assume that filament maintains a stronger state of binding when the surface is in close contact with the wide axis of the filament, that is, the minimal energy configuration aligns the $\ev_2$ with the surface normal, $\rv$.  Thus, we consider a simple energy gain per unit length proportional to $(\ev_2 \cdot \rv)^2$,
\begin{equation}
\label{eq: Eint}
E_{\rm int} = \int ds V \sin^2(\psi) ,
\end{equation}
where $V$ describes the binding affinity difference (per unit length) between ``face on" and ``edge on" binding of the filament. We note that similar models have been applied to study other aspects of biopolymer behavior, such a base-pair stacking interactions between bound pairs of DNA~\cite{haijun1,haijun2}. However, to our knowledge, our study is the first to use an intrinsically-twisted elastic filament model to study the frustrations between surface-interactions and mechanics of helical polymers, and the first to study structural transitions driven by these interactions. 

In the remainder of this article, we focus on the case of a filament bound to a planar surface ($r \to \infty$).  The lack of surface curvature greatly simplifies the coupling between $\theta$ and $\psi$,
\begin{equation}
\label{eq: Eplanar}
E(r \to \infty) =  \int ds \bigg[ \Big(\frac{C_1}{2} \cos^2 \psi + \frac{C_2}{2} \sin^2 \psi \Big) |\theta'|^2 + \frac{K}{2} (\psi'- \omega_0)^2 +V \sin^2\psi \Big] .
\end{equation}
However, due to the anisotropy of bending response, an important coupling remains between in-plane backbone fluctuations and rotations of cross section.  It is easier to bend the filament about the thin axis, and hence, rotation of $\psi$ along the backbone correlates to variations stiffness to bending in the plane.  We find below this coupling the exists significant consequences for the sensitivity of contour fluctuations of the filament backbone to the state of filament binding. We now describe our Monte Carlo simulations that serve as an independent test of our analytic results.

\subsection{Numerical Simulations}
\label{sec:sims}
In this section we describe our numerical setup to study an adsorbed filament on a flat surface. In the subsequent section, we will study the analytical solution to the equation of mechanical equilibrium for absorbed helical filaments.  While this provides an exact description of the minimal-energy states of the model, it neglects thermally-induced fluctuations of the torsion as well as the backbone bending.  It is reasonable to expect that at finite temperature, the presence of both types of fluctuations modify the ``zero-temperature" analysis of the theory.  Thus, we carry out our numerical simulations of the model to quantify the contributions of finite-temperature fluctuations absent from the analytic solutions of the theory. Monte Carlo simulations were performed to find equilibrium configurations of a $2d$ chain of length $L$ that is parsed into $N$ discrete segments or plaquettes. Each plaquette is of unit length and in general can undergo bending and torsional deformations. Plaquette $i$ has associated with it a pair of angles ($\theta_i$, $\psi_i$), which represent the local orientation and twist of the chain respectively. A discrete version of the energy given by eq. (\ref{eq: Eplanar}) is implemented to evaluate the energy of each chain configuration,

\begin{equation*}
E=\frac{1}{2}\sum_{i}^{N}\big(C_1 \cos^2(\bar{\psi}_{i,i+1}) + C_2\sin^2(\bar{\psi}_{i,i+1})\big)\big(\Delta \theta_{i,i+1}\big)^2 + K(\Delta \psi_{i,i+1}-\omega_0)^2 + V\sin^2(\psi_i).
\end{equation*}

\noindent Here $i$ refers to the segment index, $\bar{\psi}_{i,i+1}$ refers to the average twist between nearest neighbor plaquettes, $\Delta$ refers to finite differences between neighbors along the chain and elastic parameters are in units of $k_B T$. Configurations of the chain were accepted (rejected) upon comparing the trial configuration energy to the previous configuration energy using a \textsc{METROPOLIS} algorithm with the usual Boltzmann weight ($\exp(-\Delta E)$)~\cite{frenkel}.  Specifically we look for configurations $z^{(t+1)}$, where $t$ denotes the Monte Carlo step, that represent equilibrium configurations of the chain. Trial configurations of the chain are accepted/reject by the following criterion,

\[
z^{(t+1)} = \left\{ \begin{array}{ll}
z^\prime & \mbox{\text{with probability} $r(z,z^\prime) = \text{min}\big(1,\frac{p(z^\prime)}{p(z)}\big)$} \\
z & \mbox{if $r_0 > r(z,z^\prime)$} \\
\end{array}
\right. \\
.\]

\noindent Here $z$ is the previous configuration at step $t$  and $p(z^\prime)$ is the Boltzmann weight of the trial configuration at step $t+1$. Also, $r_0$ is a random number chosen from a uniform distribution between $0$ and $1$. Updates of either degree of freedom ($\theta_i$ or $\psi_i$) per plaquette were done independently with equal probability for each Monte Carlo step. Update acceptance for trail configurations of the chain were between $30\%$ and $40\%$ of the total amount of Monte Carlo steps ($10^7$ steps per chain). Data was averaged over $\sim 10^5$ equilibrium configurations of the chain. Equilibration of the chain was determined by measuring the rotational diffusion of the end to end vector using the condition,

\begin{equation*}
\vec{R}(t) \cdot\vec{R}(0)=0,
\end{equation*} 

where $t$ is the Monte Carlo step. Usual equilibration times were of the order $10^6$ steps. Chain lengths of $N=32$ were used in all numerical results unless otherwise stated.

\subsection{Thermodynamics of surface-induced untwisting}
\label{sec:twist}
In this section, we analyze the thermodynamics of surface binding of the helical filament, initially ignoring the coupling between in-plane orientation and filament twist described in eq. (\ref{eq: Eplanar}).  This can be justified in the limit of stiff filaments, where stiffness is characterized by the 2D persistence length, 
\begin{equation}
\label{eqn: leff12}
\bar{\ell}_p = 2 \frac{ \sqrt{C_1 C_2} }{k_B T} ,
\end{equation}
which is the effective persistence length of a planar helical filament twisting at a constant rate~\footnote{Notice this persistence length differs by a factor of 2 from the 3D persistence length due to the surface confinement \cite{rivetti}} In the limit $L \ll \bar{\ell}_p$, the backbone remains effectively rod-like, and we may assume $\theta' \simeq 0$.

With this approximation in mind, the remaining degrees of freedom describe filament twist, and the contact between the wide edge of the filament and surface, the final two terms in eq. (\ref{eq: Eplanar}).  The underlying frustration of helical filament binding is straightforward to understand from these terms.  While the twist elastic energy is minimized by a constant rate of rotation, $\psi = \omega_0 s$, the binding energy is minimized by locking-in to constant orientations, $\psi = n \pi$.  The equation of motion describing the minimal energy configurations of $\psi(s)$, is well-known to the analyses of 1D models of incommensurate solids~\cite{bak} and rigid pendula,
\begin{equation}
\label{eq: eom}
|\psi'|^2 = \lambda^{-2} \big( \sin^2 \psi + \epsilon\big) ,
\end{equation}
where $\epsilon$ is a positive-valued ``constant of motion" and 
\begin{equation}
\lambda^2 = \frac{K}{2 V}
\end{equation}
is a length scale defined by the ratio of twist modulus to strength of surface binding.  The limit, $\epsilon \to  0$, describes ``lock in" to an untwisted state of face-on binding, $\psi = n \pi$.  For small, but finite, values of $\epsilon$, the solution becomes inhomogeneous.  The filament locks into near perfect surface registry, with $\psi \simeq n \pi$, over spans of arc length much greater than $\lambda$.  These nearly ``commensurate" domains are separated by rapid jumps, {\it twist walls}, where $\psi$ rapidly jumps by $\pi$ over an arc-distance of order $\lambda$.  For large $\epsilon$, the filament twists homogeneously.

The transition from locked-in, untwisted filaments, to helical filaments has a highly non-linear, and critical, sensitivity to $V$ and the elastic cost (per unit length) to untwist the filament, proportional to $K \omega_0^2$.  To analyze this, we follow the analysis of the Emery and Bak~\cite{emery} to derive an effective theory for filament binding in terms of the length-averaged rate of twist, $\langle \psi ' \rangle = 2 \pi/ P$, where $P$ is the pitch of rotation.  This pitch is twice the distance between twist walls and can be computed from eq. (\ref{eq: eom}) as function of $\epsilon$,
\begin{equation}
\label{eq: P}
P( \epsilon) =  \lambda \int_0^{2\pi} \frac{d \psi }{\sqrt{\sin^2 (\psi) + \epsilon}} = 4 \lambda \epsilon^{-1/2} K(-\epsilon^{-1}) ,
\end{equation}
where $K(k)$ is the complete elliptic function of the first kind.  The asymptotic behavior near the lock-in transition has the form $P(\epsilon \to 0) \simeq - 2\lambda \ln (\epsilon /16)$, from which we express the constant, as the approximate of function of pitch, $\epsilon(P) \simeq 16 e^{- P/2 \lambda}$.  Substituting the solution, eq. (\ref{eq: eom}), into eq. (\ref{eq: Eplanar}) and integrating over one helical turn, we may derive an expression for the effective elastic energy per unit length as a function of $P$,
\begin{equation}
E(P)/P \simeq \big( 8 V \lambda - 2 \pi K \omega_0 )/P+ 4 V \lambda e^{-P / 2 \lambda}/P + C_0.
\end{equation}
Note that the density of twist walls is $2/P$, so that the first term in parentheses is proportional the energy per twist wall.  The cost of the wall corresponds an increase in potential energy over a length of order $\lambda$, while a twist wall relaxes the torsional strain on the untwisted filament, leading the negative contribution to twist wall energy.  The second term can be thought of as the cost of exponentially screened repulsive interactions between twist walls.

\begin{figure}
\center \includegraphics[clip=true, trim=50 200 50 200,scale=0.5]{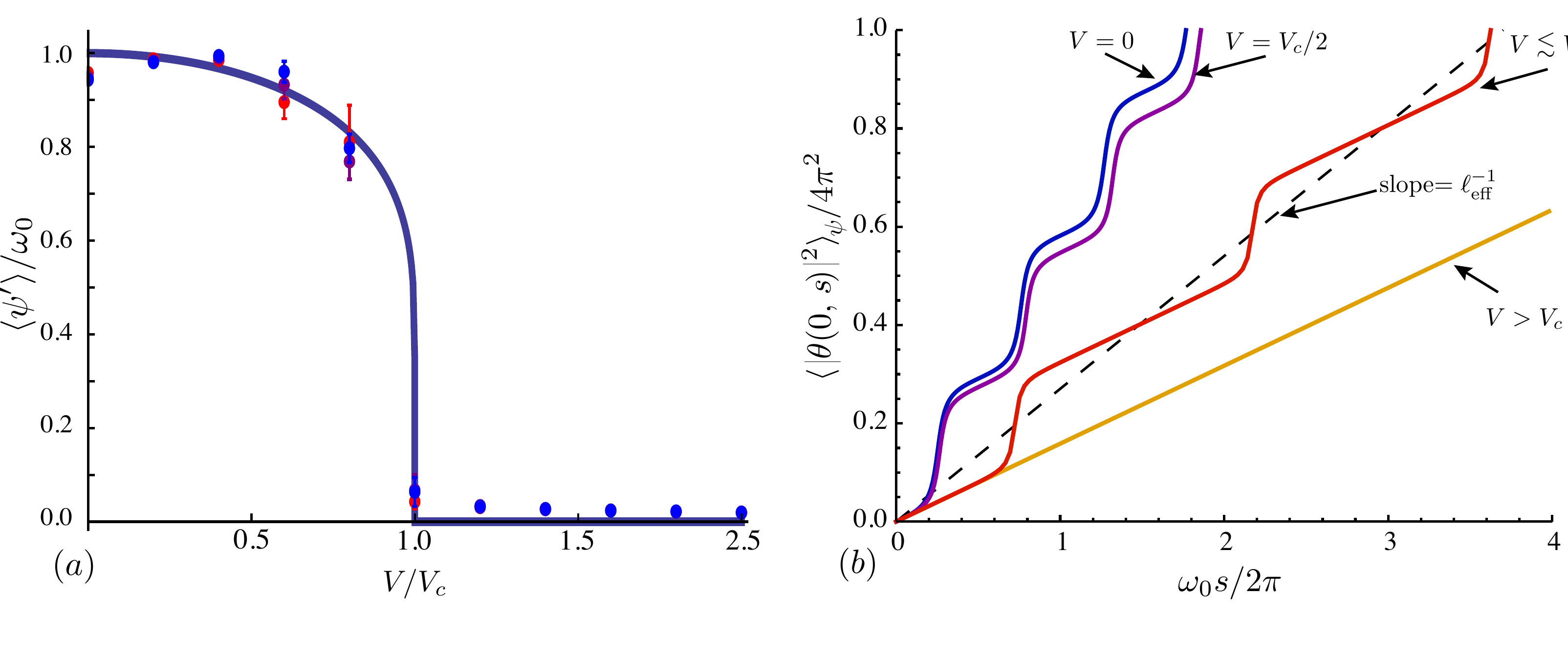}\caption{ Both figures together highlight mechanically what is taking place for both degrees of freedom as the filament becomes strongly bound to the substrate. In (a), the solid curve shows the predicted dependence of the net filament twist on the strength of surface interaction. Solid points with error bars display simulation results at finite temperature. (Red) $w/t=2.0$, (Purple) $w/t=4.0$ and (Blue) $w/t=8.0$. In (b), a plot of the angular diffusion, for $w/t=4.0$, associated in-plane orientational fluctuations (eq. (\ref{eq: diffusion}) of the filament tangent along its backbone for {\it fixed} profiles of equilibrium twist $\psi(s)$ at various levels of surface binding.  Here, $\psi(0)=0$ for each configurations.}
\label{fig: twist}
\end{figure}

It is straightforward to show that when the elastic energy gain is smaller then the potential energy cost of a twist wall, the untwisted, locked-in state ($P\to \infty$) is stable.  This occurs when $V>V_c$, where
\begin{equation}
\label{eq:Vc}
V_c = \frac{K \pi^2 \omega_0^2}{8} .
\end{equation}
However, when $V<V_c$, the energy per twist wall is negative and the filament begins unbind and wind through the incorporation of twist walls.  For small $|V_c-V|$, at the unbinding threshold, the equilibrium pitch grows rapidly as
\begin{equation}
P \sim  \ln |V_c-V |.
\end{equation}
This indicates that the filament undergoes a rapid and continuous phase transition from the untwisted and strongly-bound state to the helical and unbound state at $V=V_c$.  The dependence of the mean-twist, $2\pi/P$, can be calculated using results summarized in the Appendix.  The profile of $\langle \psi ' \rangle = 2 \pi/P$ vs. $V$ as calculated in absence of thermal fluctuations is shown in Fig.~\ref{fig: twist} (a).   In Fig.~\ref{fig: figure1} we show examples of torsional state of a helical filament with a rectangular cross section as the unwinding transitions is approached from below. Simulation data is plotted on top of  Fig.~\ref{fig: twist} (a) which is in excellent agreement with this unwinding transition.

\subsection{Conformational collapse of desorbing helical filament}
\label{sec:bend}
In this section, we consider the effect of the transition between the strongly-bound, untwisted state to the helical state on the backbone fluctuations of the surface confined filament.  Clearly, when the filament is strongly-bound in the face-on configuration in-plane bending, around the wide edge of the filament cross section, is suppressed.  As described above, just below the binding threshold, a low density of twist walls populate the length of the filament.  As the wide axis is normal to the surface in these regions, they act as localized weak spots for in-plane bending, whose bending stiffness is reduce by a factor of $(t/w)^2$ relative to the face-on spans.  Hence, in the weakly-bound, helical state, the bending stiffness becomes highly heterogeneous.

This for sufficiently anisotropic filament cross sections, this heterogeneity will have a profound impact on the bending fluctuations of the filament backbone in the plane.  These fluctuations are observable in the tangent-tangent correlations,
\begin{equation}
\langle \tv(s_1) \cdot \tv(s_2) \rangle_{\psi} = \exp \big[-\langle |\theta(s_1,s_2)|^2 \rangle_{\psi} /2 \big] .
\end{equation}
where the notation $\langle \cdot \rangle_{\psi}$ denotes an average for a {\it fixed} twist profile, $\psi(s)$.    From the mechanical energy for bending we may derive the following simple result for the ``diffusion" of $\theta$ along the contour length,
\begin{equation}
\label{eq: diffusion}
\langle |\theta(s_1,s_2)|^2 \rangle_\psi =  2 \int_{s_1}^{s_2} ds '\frac{1}{\ell_1 \cos^2\psi(s') + \ell_2 \sin^2 \psi(s') } ,
\end{equation}
where $\ell_1 =2  C_1/k_B T$ and $\ell_2 = 2 C_2/k_B T$, are 2D persistence lengths corresponding to bending around, respective, thin and wide directions of the cross section.  The experimentally measurable tangent-tangent correlation function, $C_{\tv} (s_2-s_1)=\langle \cos [ \theta(s_2) - \theta(s_1) ] \rangle$, is obtained by averaging $\langle \tv(s_1) \cdot \tv(s_2) \rangle_{\psi} $ over all torsional states.  According to the analysis above, distinct configurations of $\psi(s)$ at a given binding energy are periodic functions of $s$ (modulo shifts by $\pi$), differing only by a uniform translation along the backbone (modulo $P/2$).   We may perform the averaging over torsional configurations by fixing the twist angle at a reference point, say $\psi(s=0) = 0$, and averaging $\langle \tv(s_0) \cdot \tv(s_0+ s) \rangle_{\psi}$ over a span $P/2$ along the backbone for a given $s$
\begin{equation}
\label{eq: Ct}
C_{\tv}(s) = \frac{2}{P} \int_0^{P/2} ds_0\langle \tv(s_0) \cdot \tv(s_0+s) \rangle_{\psi}  .
\end{equation}
Hence, the zero-energy torsional fluctuations associated with sliding the twist-wall array along the backbone distribute the weak spots uniformly along the chain and lead to translationally-invariant tangent correlations.

\begin{figure}
\center \includegraphics[clip=true, trim=60 200 20 200,scale=0.55]{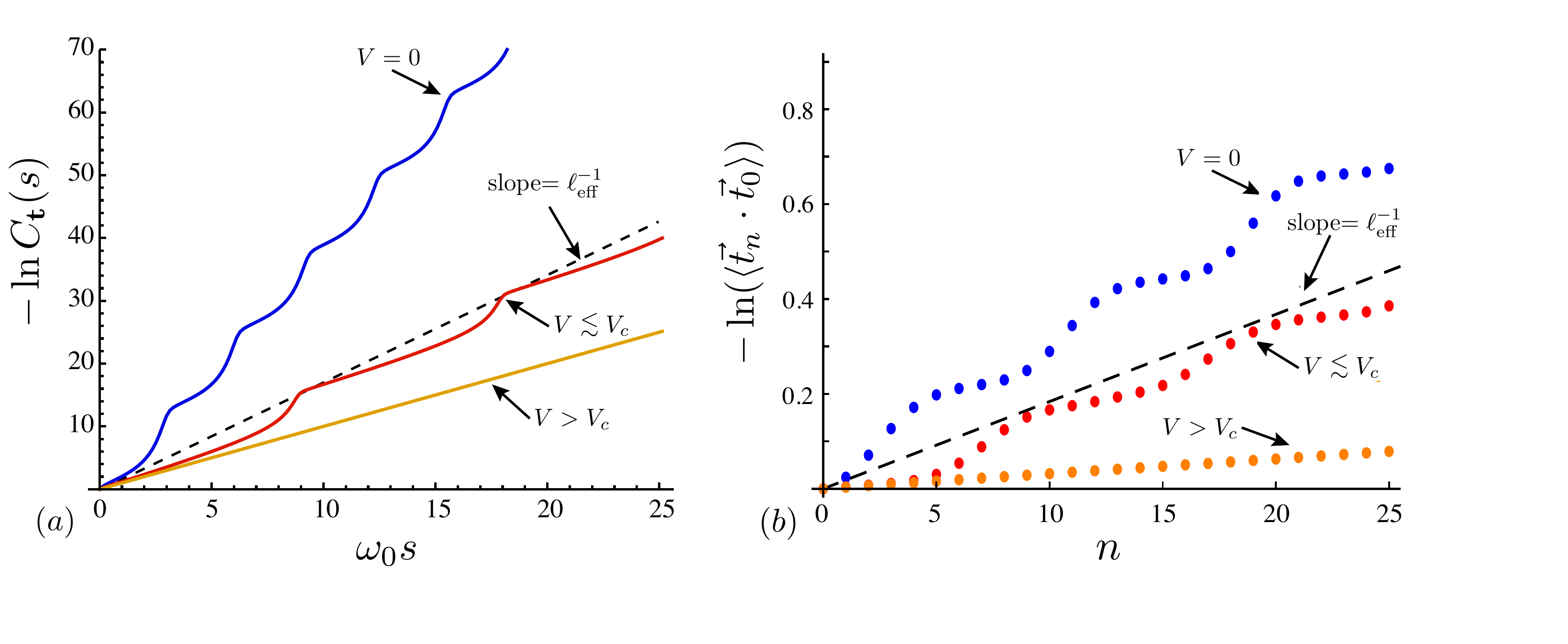}\caption{In (a), a plot of the (logarithm) of tangent-tangent correlation function of helical filament with bending anisotropy $C_1/C_2=16$ and $\omega_0 \ell_1 = 1$.   Three different surface binding strengths are shown:  (blue) vanishing binding strength; (red) near-threshold binding strength; and (yellow) above-threshold binding strength. In (b), a plot of the (logarithm) of the ensemble average tangent-tangent correlation along the chain contour $n$ for 3 different binding strengths, here $C_1/C_2=64$. As we can see the oscillations about and average effective persistence length is present in good agreement with analytical results.}
\label{fig: diffusion}
\end{figure}

In Fig.~\ref{fig: twist} (b) we plot angular excursion, $\langle |\theta(0,s)|^2 \rangle_\psi$, for filaments with a bending aspect ratio, $C_1/C_2 = 16$ and $\psi(0)=0$, for weak and strong surface binding.  Above the absorption threshold, we find that angular diffusion grows linearly with arc-length, with the relative small slope of $1/\ell_1$.  Just below the critical binding strength, the signature of twist walls appear, where the local persistence length drops to $\ell_2=\ell_1/4$ over a relativity short length span of roughly $\lambda$, corresponding to ``weak spots" of easy bending around the thin direction.  As the binding strength drops to negligible levels, the filament adopts its native twist, $\omega_0$, and the local persistence length, as measured by the inverse slope of $\langle |\theta(0,s)|^2 \rangle$, oscillates between $\ell_1$ and $\ell_2$ over equivalent quarter-pitch spans of arc-length.  Fig.~\ref{fig: diffusion} (a) shows $-\ln C_{\tv} (s)$ for a range of binding strengths, for the case $\ell_1 \omega =1$ and $C_1/C_2=16$ as in Fig.~\ref{fig: twist} (b).  Despite the translation invariance of $C_{\tv} (s)$ regularly spaced intervals of relatively flexible backbone are still apparent in the tangent-tangent correlation functions below the threshold binding strength.  

\begin{figure}
\center \includegraphics[clip=true, trim=150 240 100 200,scale=0.6]{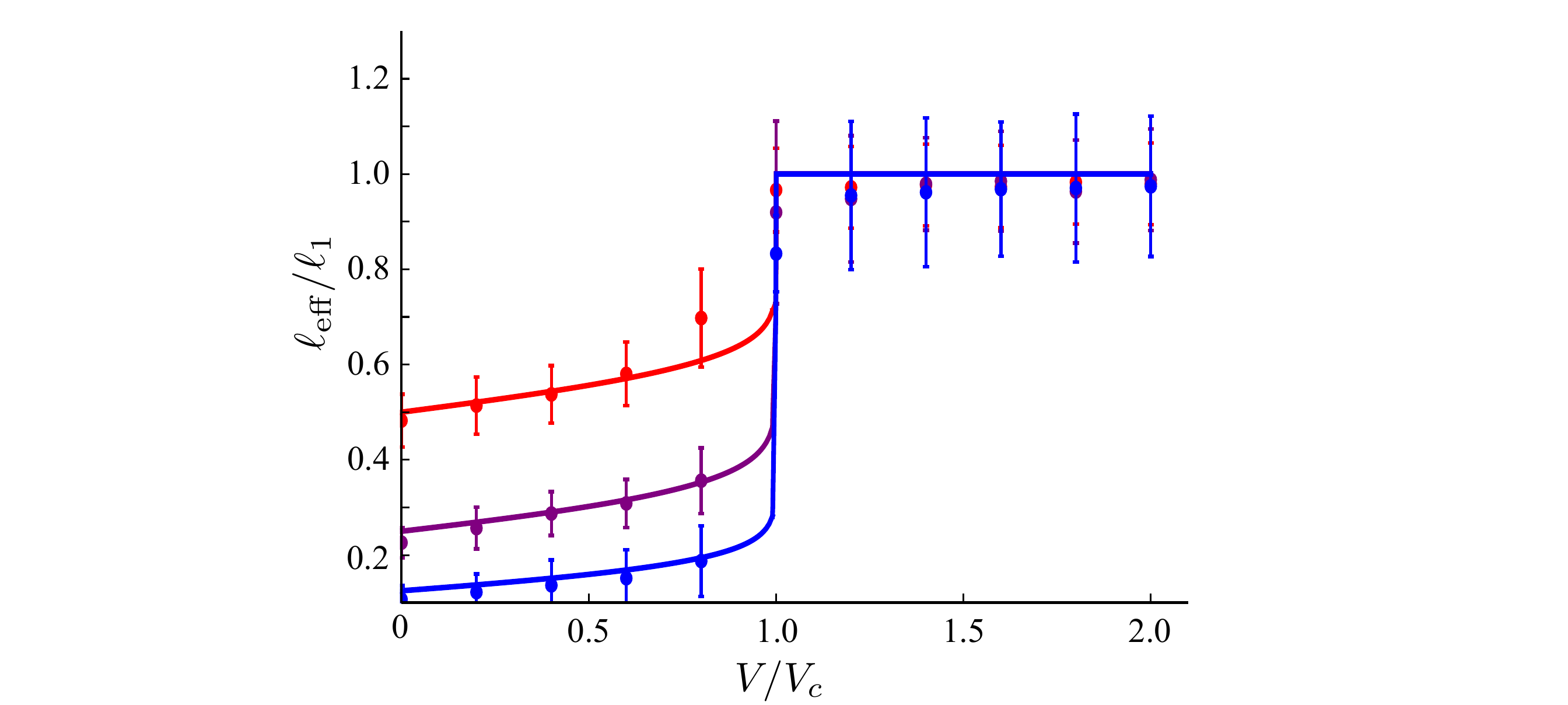}\caption{Plots of the effective persistence lengths as functions of surface binding strength for filaments of different in-plane anisotropy.  For a continuum model of elasticity, a given width and thickness ratio implies a ratio of bending moduli, $C_1= (w/t)^2 C_2$. Monte Carlo simulation data with vertical error bars are plotted on top of analytical result and show good agreement with theoretical predictions.}
\label{fig: persistence}
\end{figure}

Though the angular diffusion is not strictly a linear of function of arc-length over short distances, we may extract the long-distance, effective persistence length from the fall of tangent-tangent correlations over a span of $P/2$ corresponding to a period span of the arc-length modulation,
\begin{equation} 
\label{eq: elleff}
\ell_{\rm eff}^{-1} = -\frac{\ln C_{\tv} (P/2)}{ P/2} =  \langle|\theta(0,P/2)|^2 \rangle_\psi/P .
\end{equation}
In Fig.~\ref{fig: persistence} we plot the dependence of $\ell_{\rm eff}$ on the binding strength, for three values of cross-section anisotropy.  Clearly, in the locked-in, untwisted state, the persistence length is maximal, $\ell_{\rm eff} = \ell_1$, as in-plane bending occurs only around wide axis of the filament.  In the opposite limit of vanishing binding strength, the uniform helical rotation of the filament, $\psi = \omega_0 s$, samples bending around all cross-sectional directions evenly, and from eq. (\ref{eq: diffusion}) we find that $\ell_{\rm eff}= ( \ell_1 \ell_2)^{1/2}$.  Hence, going from strong binding to weak binding of a helical filament we observe a $(\ell_1/\ell_2)^{1/2} = w/t$-fold drop in the persistence length of the backbone.  From Fig.~\ref{fig: persistence} we see that most of this drop occurs over a narrow region near to the untwisting transition, $V \lesssim V_c$.  However, it is significant to note from Fig. \ref{fig: persistence} that even far from this transition, at weak binding strengths we predict a weaker, and roughly linear, dependence of $\ell_{\rm eff}$ on $V$.  This demonstrates that the apparent persistence length of a surface absorbed helical filament is always sensitive to anisotropic effects of surface binding. \\

\begin{figure}[h]
\center \includegraphics[clip=true, trim=150 230 100 170,scale=0.6]{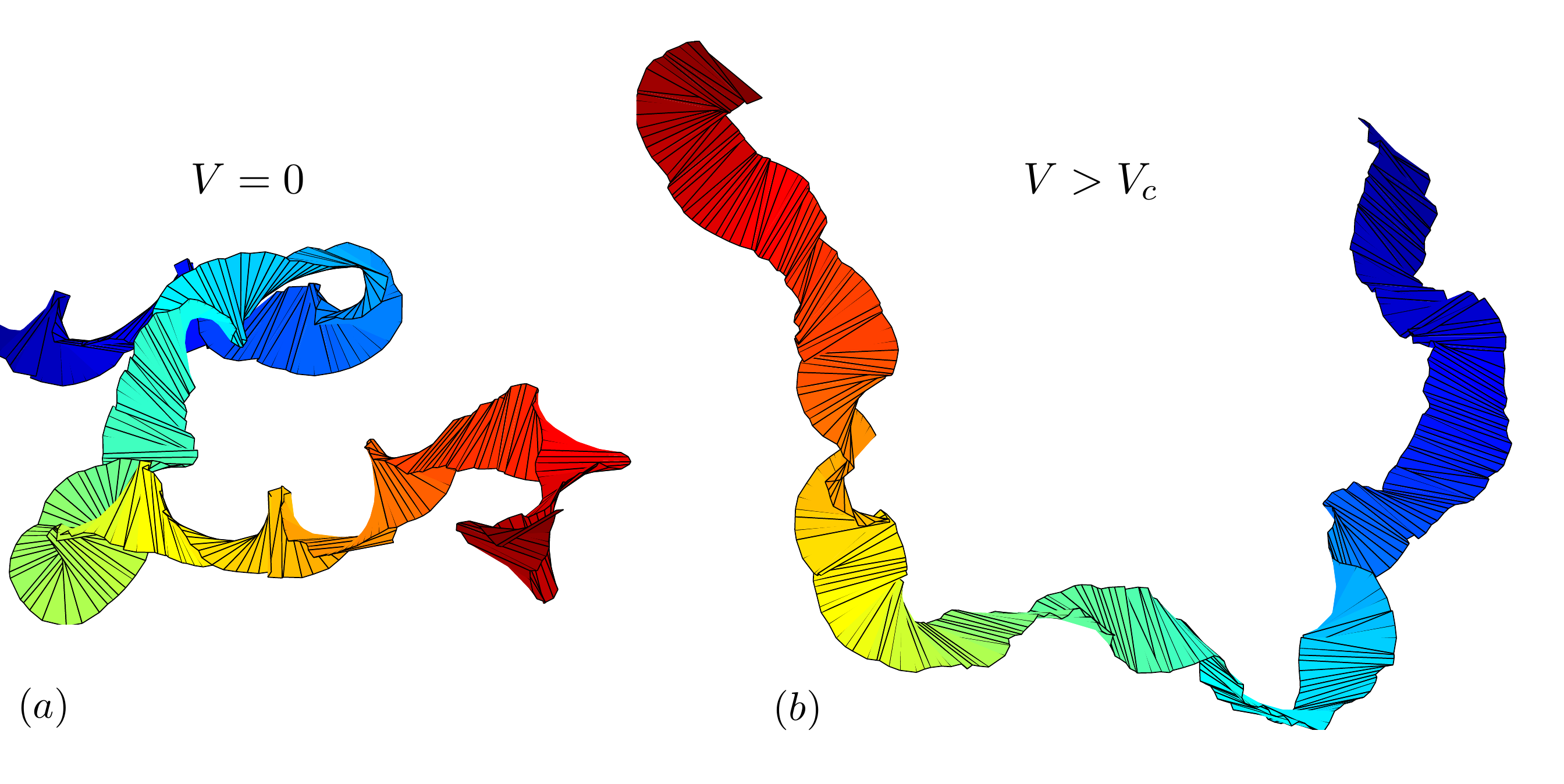}\caption{Typical Monte Carlo chain configurations acquired during simulation runtime at two values of binding potential. (a) For $V=0$ the chain twist with its natural twist rate $\omega_0$ and is collapsed with a characteristic persistence length given by eq. (\ref{eqn: leff12}). (b) For $V>V_c$ we see that twist walls are being expelled and the chain is essentially flat with $\ell_1$ as the characteristic persistence length. Here $C_1/C_2 = 2$ and L=100 subunits.} 
\label{fig: ribbons}
\end{figure}

We know look at actual conformations of the polymer as we cross this transition. Monte Carlo simulations revealed an excellent agreement with analytic results as seen in Fig.~\ref{fig: twist} and Fig.~\ref{fig: persistence}. Moreover the presence of twist walls (Fig. \ref{fig: diffusion}) can be seen in our simulation results as well.  In Fig.~\ref{fig: ribbons} we can see two chain conformations which characterize qualitatively the interplay of average twist and effective persistence length. Fig.~\ref{fig: ribbons} (a) is a snapshot of a collapsed chain configuration when $V=0$. The chain twists at its natural twist rate and the effective persistence length is leading to a collapsed chain state. In contrast Fig.~\ref{fig: ribbons} (b) shows a situation when the binding potential is above the critical potential. Here the chain is almost flat as predicted and the effective persistence length is large leading to a swollen chain configuration. 

\begin{figure}[h]
\center \includegraphics[clip=true, trim=150 240 100 100,scale=0.6]{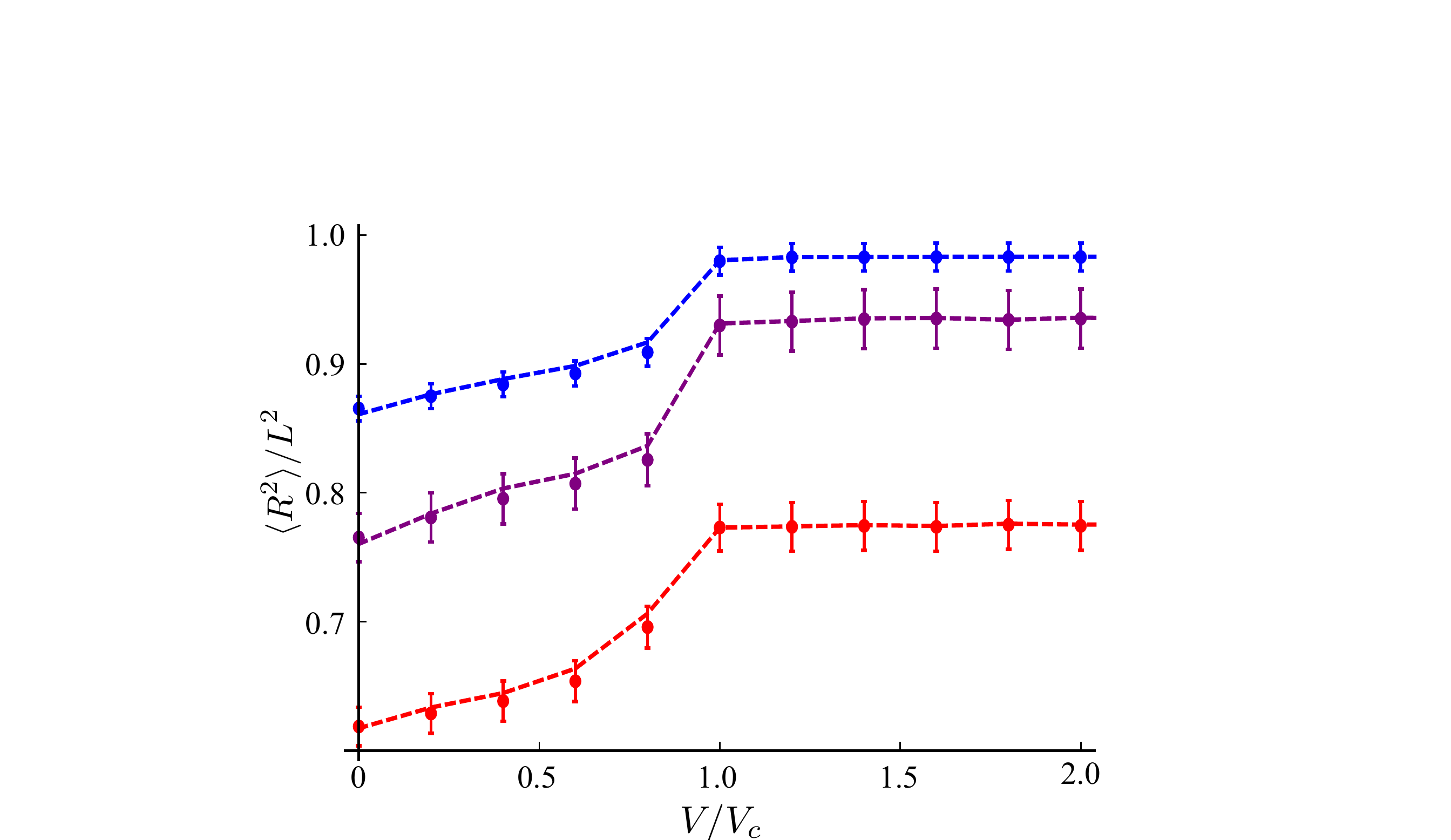} \caption{Simulation data for the end to end displacement normalized by the square of the polymer contour length. Each data set if for a fixed contour length but varying aspect ratio. As the aspect ratio of the polymer is increase from $w/t=2$ (red) to $w/t=8$ (blue) the magnitude of the transition to constant persistence length ($V>V_c$) is less dramatic. Dashed lines are generated using the $\ell_{\rm eff}$ data from figure \ref{fig: persistence} and eq. (\ref{eqn:RR2d}). } 
\label{fig: rend}
\end{figure}

To quantify the collapse transition in a more experientially accessible way, we now focus on the squared end to end displacement, $\langle R^2\rangle\rangle/L^2$ normalized by the contour length of the polymer.  For binding potentials $V<V_c$ we can see that $\langle R^2\rangle\rangle/L^2$ increases as the potential is increased from zero. This behavior should be expected from the behavior of the $\ell_{\rm eff}$ data (Fig. \ref{fig: persistence}). Once $V>V_c$ the polymer untwists and only the larger of the two bending stiffness is accessible when $\langle \psi^\prime \rangle=0$ and hence the fluctuations of the polymer should be indicative of that length scale. This picture is consistent with the data in Fig.~\ref{fig: rend} in the $V>V_c$ regime where $\langle R^2\rangle/L^2$ is constant. An expression for the squared end to end displacement for a fixed contour length in 2d is ~\cite{rivetti},

\begin{equation}
\frac{\langle R^2\rangle}{L^2} = \frac{4\ell_{p}}{L} \big( 1-\frac{2 \ell_{p}}{L}( 1-e^{( -\frac{L}{2\ell_{p}} )} \big )\big)\label{eqn:RR2d}
\end{equation} 

In Fig.~\ref{fig: rend} the dashed lines were generated using the extracted persistence lengths from the data in Fig.~\ref{fig: persistence}. As we can see the $\langle R^2\rangle/L^2$ data is in very good agreement with eq. (\ref{eqn:RR2d}).

\section{Conclusion}
\label{sec:conclusion}

Structural and mechanical properties of biopolymers adsorbed to a surface are of great importance in biology.  {\it In vivo} there are many examples of membrane bound polymers which are essential for cellular functions such as cell shape maintenance, motility, mechanical sensing and the cleaving and contraction of the cell membrane during division. It is therefore imperative to understand if and how biopolymers change their mechanical and structural properties when adsorbed/bound to a membrane surface. This question becomes particularly interesting when one considers that real biopolymers typically have helical structures and can have anisotropic bending stiffnesses. The interplay between the intrinsic helicity, the binding affinity to the surface and the bending stiffness anisotropy lead to non-trivial conformational and mechanical changes upon filament binding, which in turn can be tuned or regulated by the filament's affinity for the surface.\par

Here we have studied a continuum model for a ribbon polymer which has an anisotropic bending stiffness and a native helical twist around it's long axis. While the anisotropy in a real biopolymer can originate in the detailed chemical structure, here we choose to model the biopolymer as uniformly elastic ribbon with a width $w$ and thickness $t$, thus the structurally the anisotropic bending stiffness arises when aspect ratio $w/t \neq1$. The helical twist implies that the filament's affinity for the surface will be modulated on a period set by the native twist rate when the filament is in its naturally twisted state.  In the context of the ribbon model we assume that the wider side with more surface area also has a higher binding affinity for the surface and the twist therefore determines which side is in contact with the surface.  Though this assumption is follows intuitively from the fact that face-on binding of the wider edge of the cross-section allows for greater filament-surface contact, the detailed orientation dependence of filament-surface interactions need not conform to this simplified picture.  It is quite easy to see that, however, that any sufficiently strong orientation dependence of interactions (say, a strong preference edge-on binding) will give rise to the unwinding transition described here. \par

We first considered an infinitely stiff limit ($\theta^\prime\simeq 0$) where the larger surface area is in contact with the binding surface most of the time. For surface potential strengths above the critical value $V_c$ (eq. \ref{eq:Vc}), the net twist vanishes and the wider face of the ribbon maintains contact with the surface. Once below the critical value of the potential ($V\lesssim V_c$), a sharp continuous transition occurs to a configuration where twist walls proliferate and form a periodic array (in the absence of thermal fluctuations) which are are separated by a distance $P/2$.  Due to the highly heterogeneous structure of the twist in this transition from the weakly-bound to strongly-bound regime, we found that spectrum of in-plane bending fluctuations is highly sensitive to surface interactions near to this transition.  In the weak-binding limit, thermal bending of the backbone is enhanced in the ``softer" regions where the filament has an edge-on orientation.  Here, excursions of the in-plane angle $\theta$ oscillate around a purely diffusive dependence in the contour length, a signature of the appearance of twist walls which function as floppy joints with the period of the oscillations being set by the average distance between twist walls . The effective persistence length has a highly nonlinear increase with increased potential for $V< V_c$. As the magnitude of the potential approaches the critical potential from below we find that the amplitude of the oscillations diminishes to zero as the period continuously approaches infinity. Mechanically this is due to the larger flat side constituting more and more of the bound side and thus the effective persistence length approaches that of the stiffer bending direction, leading to a homogeneous bending stiffness corresponding to the bending about the wide filament direction.  

We propose that one could use the strong dependence of the apparent persistence length on surface interactions to address in important and previously unresolved biomechanical property of biological filaments experimentally.  Though the bending stiffness of helical biofilaments like F-actin is well characterized on long lengthscales, the are no measurements to our knowledge of the local anisotropy of bending stiffness.   By modulating the strength of filament-surface interactions and monitoring fluctuations of the backbone orientation, say by modulating surface charge or osmotic pressure of the solution, the dependence of effective persistence length on binding strength, $\ell_{\rm eff} (V)$ can be measured by standard microscopy techniques.  Extrapolating these data to the limits of low- and high-surface binding one has a direct access to stiffness anisotropy via the relation, $\ell_{\rm eff}(V \to \infty)/\ell_{\rm eff}(V \to 0) = \sqrt{C_1/C_2}$.

To assess the physiological relevance and experimental accessibility of surface-induced untwisting helical biopolymers, we now discuss known values of the physical parameters affecting this behavior.  For simplicity we focus on electrostatic interactions between the polymer and surface, which is easily accessible {\it in vitro}.  We are interested then in the surface charge density, $\sigma$, required to attain the critical binding strength $V_c$ say for F-actin whose long-wavelength mechanical properties are well-characterized.
F-actin is negatively charged with linear charge density of $\rho \simeq 4 ~{\rm nm}^{-1}$ (in units of $e$)~\cite{tang}, and thus in solution would bind to an oppositely charged surface. Given the equation for the critical binding strength (eqn.\ref{eq:Vc}) we can calculate the necessary charge density. First we quantify the critical potential per unit-length that is required to untwist F-actin, we use a native twist of 1.6 turns per micron ($10~{\rm \mu m}^{-1}$)~\cite{holmes} and a twist stiffness in the range of $8 \times 10^{-2}~{\rm pN}~{\rm \mu m}^2$ This provides us with a scale of critical potential (per-unit length), 

\begin{equation}
V_c \approx 10~{\rm pN}
\end{equation}

To quantify the critical surface charge we assume that the solution of F-actin is dilute such that the distance between neighboring adsorbed filaments is very large compared to their length.  Assuming a screened electrostatics, the electrostatic free energy gained per unit charge brought to an oppositely charged surface is roughly $k_B T\sigma  \ell_B/\kappa $, where $\ell_B$ is the Bjerrum length and $\kappa^{-1}$ is the Debye screening length of the solution.  From this we may crudely estimate the value of the surface-binding potential to be of order
\begin{equation*}
V/k_B T \approx \frac{\sigma \rho \ell_B}{\kappa}.
\end{equation*}

From this we estimate the value of $\sigma$ for which we expect surface interactions to be sufficiently strong to drive the unwinding transtion of F-actin, $\sigma \approx V_c \kappa/ k_B T \rho \ell_B$,  at room temperature in  $1\hspace*{3pt}\mu M$ monovalent salt, where $\ell_B /\kappa =0.21~{\rm nm}^2$. The critical surface charge to fully untwist the filament is,

\begin{equation*}
\sigma_c =3.e~{\rm nm}^{-2}
\end{equation*}

These values are consistent with physiological values for the surface charge density on cell membranes indicating that such conformational transitions and the associated changes in mechanical properties of surface bound filaments may in fact be exploited {\it in vivo}. The modest scale $\sigma_c$ required also suggest that such the a careful regulation of  electrostatically-induced surface binding can be used to indirectly probe the anisotropic mechanical properties of helical helical biofilaments like F-actin and bacterial homologs like MreB. It is important to take these effects into account when backing out elastic properties of twisted filaments adsorbed on surfaces. For instance, there have been recent observations that DNA can form abnormally high bending deformations when adsorbed to a surface~\cite{wiggins}. DNA is also intrinsically twisted (1 revolution per 3.4 nm) on length scales where anomalous elasticity is observed ($\sim 5$ nm). In the context of our model we have presented here it would be interesting to understand how local denatured ``weak spots" could affect the formation of twist walls and how the overall observed persistence length is determined by the interplay of intrinsic twist and adsorption onto a surface.

\begin{acknowledgments}
The authors would like to thank the Kavli Institute for Theoretical Physics (supported by NSF PHY11-25915) where some of this work was done.  GG was supported by the NSF Career program under DMR Grant 09-55760.  AG would also like to acknowledge support from a James S. McDonnell Foundation Award, NSF grant DBI-0960480, NSF grant EF-1038697, a UC MEXUS grant, and a George E. Brown, Jr. Award.
\end{acknowledgments}

\appendix

\section{Twist profile solutions}
Given the constant of integration, $\epsilon$, the relationship between $\psi$ and arc-length follows the solution to eq. (\ref{eq: eom}),
\begin{equation}
s = \lambda \int_0^\psi d \psi' \frac{1}{\sqrt{\sin^2 \psi' + \epsilon}} =  \lambda \epsilon^{-1/2} F(\psi,-\epsilon^{-1}),
\end{equation}
where $F(x, k)$ is the incomplete elliptic integral of the first kind and we have set $\psi(s=0) = 0$.  This equation can be inverted to given $\psi(s)$ in terms of the Jacobi elliptic amplitude function,
\begin{equation}
\psi(s) = {\rm am} \Big(
\frac{s\sqrt{\epsilon} }{\lambda},  -\epsilon^{-1} \Big) .
\end{equation} 
Using the solution to Euler-Lagrange equation for $\psi(s)$, we may derive an expression for the energy of a single pitch of the bound helix as function of the parameter $\epsilon$,
\begin{equation}
E(\epsilon) = \int_0^{P(\epsilon)} ds \Big[ \frac{K}{2} |\psi'|^2 + V \sin^2\psi \Big] - 2 \pi K \omega_0 + \frac{K}{2} \omega_0^2 P(\epsilon) .
\end{equation}
Using eq. (\ref{eq: eom}) we may the integral above as,
\begin{equation}
 \int_0^{P(\epsilon)} ds \Big[ \frac{K}{2} |\psi'|^2 + V \sin^2\psi \Big] = 2 V \lambda  \int_0^{2 \pi} dp \sqrt{ \sin^2 p + \epsilon} - V \epsilon P(\epsilon) .
\end{equation}
From this, we have the energy density,
\begin{equation}
\frac{ E(\epsilon) }{ P(\epsilon)} = \frac{2 V \lambda}{P(\epsilon)}  \int_0^{2 \pi} d \psi \sqrt{ \sin^2 \psi + \epsilon} - V\epsilon - \frac{2 \pi K \omega_0}{ P(\epsilon)} ,
\end{equation}
which we minimize with respect to $\epsilon$ to derive the equation of state, effectively relating mean pitch to the binding potential and twist-elastic cost for $V \leq V_c$,
\begin{equation}
\frac{ \pi K \omega_0}{  V \lambda} = \int_0^{2 \pi} d \psi \sqrt{ \sin^2 \psi + \epsilon} = 4\epsilon^{1/2} E( -\epsilon^{-1}),
\end{equation}
where $E(k)$ is the complete elliptic function of the second kind. For a given $\epsilon$ this relation can easily be solved to binding strength,
\begin{equation}
\label{eq: Vep}
V(\epsilon)=\frac{ V_c }{\epsilon^{1/2} E( -\epsilon^{-1}) } .
\end{equation}
Noting that $\lambda = \lambda_c (V/V_c)^{-1/2}$, where $\lambda_c^{-1} = \pi \omega_0 /2$, eqs. (\ref{eq: Vep}) we may rewrite the equilibrium pitch in terms of complete elliptic integrals,
\begin{equation}
P(\epsilon) = \lambda_c E( -\epsilon^{-1}) K( -\epsilon^{-1}) .
\end{equation}
Using these expressions, $V(\epsilon)$ and $P(\epsilon)$, we describe the relationship between equilibrium pitch of the helical filament and binding strength through the parametric dependence on $\epsilon$, plotted in Fig.~\ref{fig: twist} (b).

\bibliography{ref} %your .bib file

\providecommand*{\mcitethebibliography}{\thebibliography}
\csname @ifundefined\endcsname{endmcitethebibliography}
{\let\endmcitethebibliography\endthebibliography}{}
\begin{mcitethebibliography}{39}
\providecommand*{\natexlab}[1]{#1}
\providecommand*{\mciteSetBstSublistMode}[1]{}
\providecommand*{\mciteSetBstMaxWidthForm}[2]{}
\providecommand*{\mciteBstWouldAddEndPuncttrue}
  {\def\EndOfBibitem{\unskip.}}
\providecommand*{\mciteBstWouldAddEndPunctfalse}
  {\let\EndOfBibitem\relax}
\providecommand*{\mciteSetBstMidEndSepPunct}[3]{}
\providecommand*{\mciteSetBstSublistLabelBeginEnd}[3]{}
\providecommand*{\EndOfBibitem}{}
\mciteSetBstSublistMode{f}
\mciteSetBstMaxWidthForm{subitem}
{(\emph{\alph{mcitesubitemcount}})}
\mciteSetBstSublistLabelBeginEnd{\mcitemaxwidthsubitemform\space}
{\relax}{\relax}

\bibitem[Lodish and et~al.(1995)]{lodish}
H.~Lodish and et~al., \emph{Molecular Cell Biology}, W.H. Freeman,New York, 3rd
  edn, 1995\relax
\mciteBstWouldAddEndPuncttrue
\mciteSetBstMidEndSepPunct{\mcitedefaultmidpunct}
{\mcitedefaultendpunct}{\mcitedefaultseppunct}\relax
\EndOfBibitem
\bibitem[Cowin and Burke(1996)]{cowin}
P.~Cowin and B.~Burke, \emph{Curr.Op.Cell Biol.}, 1996, \textbf{8},
  56--65.\relax
\mciteBstWouldAddEndPunctfalse
\mciteSetBstMidEndSepPunct{\mcitedefaultmidpunct}
{}{\mcitedefaultseppunct}\relax
\EndOfBibitem
\bibitem[Drewes \emph{et~al.}(1998)Drewes, Ebneth, and Mandelkow]{mts}
G.~Drewes, A.~Ebneth and E.~Mandelkow, \emph{Trends Biochem. Sci.}, 1998,
  \textbf{23}, 30717\relax
\mciteBstWouldAddEndPuncttrue
\mciteSetBstMidEndSepPunct{\mcitedefaultmidpunct}
{\mcitedefaultendpunct}{\mcitedefaultseppunct}\relax
\EndOfBibitem
\bibitem[Lipowsky and Sackmann(1995)]{janmey}
R.~Lipowsky and E.~Sackmann, \emph{Handbook of Biological Physics, 1: The
  Structure and dynamics of membranes. (Chapter 17)}, Amsterdam: Elsevier
  Science, Hoff A (Ed) edn, 1995\relax
\mciteBstWouldAddEndPuncttrue
\mciteSetBstMidEndSepPunct{\mcitedefaultmidpunct}
{\mcitedefaultendpunct}{\mcitedefaultseppunct}\relax
\EndOfBibitem
\bibitem[Stuurman \emph{et~al.}(1998)Stuurman, Heins, and Aebi]{lamin}
N.~Stuurman, S.~Heins and U.~Aebi, \emph{Journal of Structural Biology}, 1998,
  \textbf{122}, 42--46\relax
\mciteBstWouldAddEndPuncttrue
\mciteSetBstMidEndSepPunct{\mcitedefaultmidpunct}
{\mcitedefaultendpunct}{\mcitedefaultseppunct}\relax
\EndOfBibitem
\bibitem[Kamasaki \emph{et~al.}(2007)Kamasaki, Osumi, and Mabuchi]{cont}
T.~Kamasaki, M.~Osumi and I.~Mabuchi, \emph{J. Cell Biol.}, 2007, \textbf{178},
  178:765171\relax
\mciteBstWouldAddEndPuncttrue
\mciteSetBstMidEndSepPunct{\mcitedefaultmidpunct}
{\mcitedefaultendpunct}{\mcitedefaultseppunct}\relax
\EndOfBibitem
\bibitem[Lutkenhaus(1993)]{ftsz}
J.~Lutkenhaus, \emph{Molecular Microbiolog}, 1993, \textbf{9}, 403--9.\relax
\mciteBstWouldAddEndPunctfalse
\mciteSetBstMidEndSepPunct{\mcitedefaultmidpunct}
{}{\mcitedefaultseppunct}\relax
\EndOfBibitem
\bibitem[Jensen \emph{et~al.}(2005)Jensen, Thompson, and Harry]{ftsa}
S.~Jensen, L.~Thompson and E.~Harry, \emph{J Bacteriol}, 2005, \textbf{187},
  65361744\relax
\mciteBstWouldAddEndPuncttrue
\mciteSetBstMidEndSepPunct{\mcitedefaultmidpunct}
{\mcitedefaultendpunct}{\mcitedefaultseppunct}\relax
\EndOfBibitem
\bibitem[Andrews and Arkin.(2007)]{arkin}
S.~Andrews and A.~P. Arkin., \emph{Biophys. J.}, 2007, \textbf{93},
  1872--1884\relax
\mciteBstWouldAddEndPuncttrue
\mciteSetBstMidEndSepPunct{\mcitedefaultmidpunct}
{\mcitedefaultendpunct}{\mcitedefaultseppunct}\relax
\EndOfBibitem
\bibitem[van~den Ent \emph{et~al.}(2010)van~den Ent, Johnson, Persons, de~Boer,
  and Lwe]{mreb}
F.~van~den Ent, C.~Johnson, L.~Persons, P.~de~Boer and J.~Lwe, \emph{EMBO J},
  2010, \textbf{29}, 10811790\relax
\mciteBstWouldAddEndPuncttrue
\mciteSetBstMidEndSepPunct{\mcitedefaultmidpunct}
{\mcitedefaultendpunct}{\mcitedefaultseppunct}\relax
\EndOfBibitem
\bibitem[Salje and et~al.(2011)]{mrebdir}
J.~Salje and et~al., \emph{Mol. Cell}, 2011, \textbf{43}, 478177\relax
\mciteBstWouldAddEndPuncttrue
\mciteSetBstMidEndSepPunct{\mcitedefaultmidpunct}
{\mcitedefaultendpunct}{\mcitedefaultseppunct}\relax
\EndOfBibitem
\bibitem[Fleer \emph{et~al.}(1993)Fleer, Cohen-Stuart, Scheutjens, Cosgrove,
  and Vincent]{polyint}
G.~Fleer, M.~Cohen-Stuart, J.~Scheutjens, T.~Cosgrove and B.~Vincent,
  \emph{Polymers at Interfaces}, Chapman and Hall, London, 1993\relax
\mciteBstWouldAddEndPuncttrue
\mciteSetBstMidEndSepPunct{\mcitedefaultmidpunct}
{\mcitedefaultendpunct}{\mcitedefaultseppunct}\relax
\EndOfBibitem
\bibitem[Eisenriegler(1993)]{polyint2}
E.~Eisenriegler, \emph{Polymers at Interfaces}, World Scientific, Singapore,
  1993\relax
\mciteBstWouldAddEndPuncttrue
\mciteSetBstMidEndSepPunct{\mcitedefaultmidpunct}
{\mcitedefaultendpunct}{\mcitedefaultseppunct}\relax
\EndOfBibitem
\bibitem[Shin and Grason(2010)]{greg1}
H.~Shin and G.~M. Grason, \emph{Phys. Rev. E}, 2010, \textbf{82}, 051919\relax
\mciteBstWouldAddEndPuncttrue
\mciteSetBstMidEndSepPunct{\mcitedefaultmidpunct}
{\mcitedefaultendpunct}{\mcitedefaultseppunct}\relax
\EndOfBibitem
\bibitem[Shin \emph{et~al.}(2009)Shin, Purdy, James, Bartles, Wong, and
  Grason]{greg2}
H.~Shin, K.~R. Purdy, D.~James, R.~Bartles, G.~C.~L. Wong and G.~M. Grason,
  \emph{Phys. Rev. Lett.}, 2009, \textbf{103}, 238102\relax
\mciteBstWouldAddEndPuncttrue
\mciteSetBstMidEndSepPunct{\mcitedefaultmidpunct}
{\mcitedefaultendpunct}{\mcitedefaultseppunct}\relax
\EndOfBibitem
\bibitem[Grason(2009)]{greg3}
G.~M. Grason, \emph{Phys. Rev. E}, 2009, \textbf{79}, 041919\relax
\mciteBstWouldAddEndPuncttrue
\mciteSetBstMidEndSepPunct{\mcitedefaultmidpunct}
{\mcitedefaultendpunct}{\mcitedefaultseppunct}\relax
\EndOfBibitem
\bibitem[Grason and Bruinsma(2007)]{greg4}
G.~M. Grason and R.~F. Bruinsma, \emph{Phys. Rev. Lett.}, 2007, \textbf{99},
  098101\relax
\mciteBstWouldAddEndPuncttrue
\mciteSetBstMidEndSepPunct{\mcitedefaultmidpunct}
{\mcitedefaultendpunct}{\mcitedefaultseppunct}\relax
\EndOfBibitem
\bibitem[Heussinger \emph{et~al.}(2010)Heussinger, Schller, and Frey]{1rib1}
C.~Heussinger, F.~Schller and E.~Frey, \emph{Phys. Rev. E}, 2010, \textbf{81},
  021904\relax
\mciteBstWouldAddEndPuncttrue
\mciteSetBstMidEndSepPunct{\mcitedefaultmidpunct}
{\mcitedefaultendpunct}{\mcitedefaultseppunct}\relax
\EndOfBibitem
\bibitem[Alim and Frey(24)]{1rib2}
K.~Alim and E.~Frey, \emph{The European Physical Journal E}, 24, \textbf{185},
  2007\relax
\mciteBstWouldAddEndPuncttrue
\mciteSetBstMidEndSepPunct{\mcitedefaultmidpunct}
{\mcitedefaultendpunct}{\mcitedefaultseppunct}\relax
\EndOfBibitem
\bibitem[Rappaport and Rabin(2007)]{1rib3}
S.~M. Rappaport and Y.~Rabin, \emph{Journal of Physics A Mathematical and
  Theoretical}, 2007, \textbf{40}, 4455\relax
\mciteBstWouldAddEndPuncttrue
\mciteSetBstMidEndSepPunct{\mcitedefaultmidpunct}
{\mcitedefaultendpunct}{\mcitedefaultseppunct}\relax
\EndOfBibitem
\bibitem[Arinstein(2005)]{1rib4}
A.~E. Arinstein, \emph{Phys. Rev. E}, 2005, \textbf{72}, 051805\relax
\mciteBstWouldAddEndPuncttrue
\mciteSetBstMidEndSepPunct{\mcitedefaultmidpunct}
{\mcitedefaultendpunct}{\mcitedefaultseppunct}\relax
\EndOfBibitem
\bibitem[Col and Liverpool(2004)]{1rib5}
A.~D. Col and T.~B. Liverpool, \emph{Phys. Rev. E}, 2004, \textbf{69},
  061907\relax
\mciteBstWouldAddEndPuncttrue
\mciteSetBstMidEndSepPunct{\mcitedefaultmidpunct}
{\mcitedefaultendpunct}{\mcitedefaultseppunct}\relax
\EndOfBibitem
\bibitem[Mergell \emph{et~al.}(2002)Mergell, Ejtehadi, and Everaers]{1rib6}
B.~Mergell, M.~R. Ejtehadi and R.~Everaers, \emph{Phys. Rev. E}, 2002,
  \textbf{66}, 011903\relax
\mciteBstWouldAddEndPuncttrue
\mciteSetBstMidEndSepPunct{\mcitedefaultmidpunct}
{\mcitedefaultendpunct}{\mcitedefaultseppunct}\relax
\EndOfBibitem
\bibitem[Bouchiat and Mezard(1998)]{1rib7}
C.~Bouchiat and M.~Mezard, \emph{Phys. Rev. Lett.}, 1998, \textbf{80},
  1556\relax
\mciteBstWouldAddEndPuncttrue
\mciteSetBstMidEndSepPunct{\mcitedefaultmidpunct}
{\mcitedefaultendpunct}{\mcitedefaultseppunct}\relax
\EndOfBibitem
\bibitem[Panyukov and Rabin(2000)]{1rib8}
S.~Panyukov and Y.~Rabin, \emph{Phys. Rev. Lett.}, 2000, \textbf{85},
  2404\relax
\mciteBstWouldAddEndPuncttrue
\mciteSetBstMidEndSepPunct{\mcitedefaultmidpunct}
{\mcitedefaultendpunct}{\mcitedefaultseppunct}\relax
\EndOfBibitem
\bibitem[Liverpool \emph{et~al.}(1998)Liverpool, Golestanian, and
  Kremer]{1rib9}
T.~Liverpool, R.~Golestanian and K.~Kremer, \emph{Phys. Rev. Lett}, 1998,
  \textbf{80}, 40\relax
\mciteBstWouldAddEndPuncttrue
\mciteSetBstMidEndSepPunct{\mcitedefaultmidpunct}
{\mcitedefaultendpunct}{\mcitedefaultseppunct}\relax
\EndOfBibitem
\bibitem[Golestanian and Liverpool(2000)]{1rib10}
R.~Golestanian and T.~Liverpool, \emph{Phys. Rev. E}, 2000, \textbf{62},
  5488\relax
\mciteBstWouldAddEndPuncttrue
\mciteSetBstMidEndSepPunct{\mcitedefaultmidpunct}
{\mcitedefaultendpunct}{\mcitedefaultseppunct}\relax
\EndOfBibitem
\bibitem[Nyrkova \emph{et~al.}(1996)Nyrkova, Semenov, Joanny, and
  Khokhlov]{1rib11}
I.~A. Nyrkova, A.~N. Semenov, J.-F. Joanny and A.~R. Khokhlov, \emph{J. Phys.
  II}, 1996, \textbf{6}, 1411\relax
\mciteBstWouldAddEndPuncttrue
\mciteSetBstMidEndSepPunct{\mcitedefaultmidpunct}
{\mcitedefaultendpunct}{\mcitedefaultseppunct}\relax
\EndOfBibitem
\bibitem[Giomi and Mahadevan(2010)]{2rib1}
L.~Giomi and L.~Mahadevan, \emph{Phys. Rev. Lett.}, 2010, \textbf{104},
  238104\relax
\mciteBstWouldAddEndPuncttrue
\mciteSetBstMidEndSepPunct{\mcitedefaultmidpunct}
{\mcitedefaultendpunct}{\mcitedefaultseppunct}\relax
\EndOfBibitem
\bibitem[Suzuki and Williams(2009)]{2rib2}
Y.~Y. Suzuki and D.~R.~M. Williams, \emph{EPL}, 2009, \textbf{85}, 63001\relax
\mciteBstWouldAddEndPuncttrue
\mciteSetBstMidEndSepPunct{\mcitedefaultmidpunct}
{\mcitedefaultendpunct}{\mcitedefaultseppunct}\relax
\EndOfBibitem
\bibitem[Landau and Lifshitz(1986)]{landau}
L.~D. Landau and E.~M. Lifshitz, \emph{Theory of Elasticity}, Pergamon, Oxford,
  3rd edn, 1986\relax
\mciteBstWouldAddEndPuncttrue
\mciteSetBstMidEndSepPunct{\mcitedefaultmidpunct}
{\mcitedefaultendpunct}{\mcitedefaultseppunct}\relax
\EndOfBibitem
\bibitem[Haijun \emph{et~al.}(1999)Haijun, Yang, and Zhong-can]{haijun1}
Z.~Haijun, Z.~Yang and O.-Y. Zhong-can, \emph{PRL}, 1999, \textbf{82},
  4650--3\relax
\mciteBstWouldAddEndPuncttrue
\mciteSetBstMidEndSepPunct{\mcitedefaultmidpunct}
{\mcitedefaultendpunct}{\mcitedefaultseppunct}\relax
\EndOfBibitem
\bibitem[Haijun \emph{et~al.}(2000)Haijun, Yang, and Zhong-can]{haijun2}
Z.~Haijun, Z.~Yang and O.-Y. Zhong-can, \emph{PRE.}, 2000, \textbf{62},
  1045--58\relax
\mciteBstWouldAddEndPuncttrue
\mciteSetBstMidEndSepPunct{\mcitedefaultmidpunct}
{\mcitedefaultendpunct}{\mcitedefaultseppunct}\relax
\EndOfBibitem
\bibitem[Bak(1982)]{bak}
P.~Bak, \emph{Rep. Prog. Phys.}, 1982, \textbf{45}, 557\relax
\mciteBstWouldAddEndPuncttrue
\mciteSetBstMidEndSepPunct{\mcitedefaultmidpunct}
{\mcitedefaultendpunct}{\mcitedefaultseppunct}\relax
\EndOfBibitem
\bibitem[Bak and Emery(1976)]{emery}
P.~Bak and V.~J. Emery, \emph{Phys. Rev. Lett.}, 1976, \textbf{36}, 978\relax
\mciteBstWouldAddEndPuncttrue
\mciteSetBstMidEndSepPunct{\mcitedefaultmidpunct}
{\mcitedefaultendpunct}{\mcitedefaultseppunct}\relax
\EndOfBibitem
\bibitem[Rivetti \emph{et~al.}(1996)Rivetti, Guthold, and Bustamante]{rivetti}
C.~Rivetti, M.~Guthold and C.~Bustamante, \emph{J. Mol. Biol.}, 1996,
  \textbf{264}, 919\relax
\mciteBstWouldAddEndPuncttrue
\mciteSetBstMidEndSepPunct{\mcitedefaultmidpunct}
{\mcitedefaultendpunct}{\mcitedefaultseppunct}\relax
\EndOfBibitem
\bibitem[Tang and Janmey(1996)]{tang}
J.~X. Tang and P.~A. Janmey, \emph{J. Biol. Chem.}, 1996, \textbf{271},
  8556\relax
\mciteBstWouldAddEndPuncttrue
\mciteSetBstMidEndSepPunct{\mcitedefaultmidpunct}
{\mcitedefaultendpunct}{\mcitedefaultseppunct}\relax
\EndOfBibitem
\bibitem[Holmes \emph{et~al.}(1990)Holmes, Popp, Gebhard, and Kabsch]{holmes}
K.~C. Holmes, D.~Popp, W.~Gebhard and W.~Kabsch, \emph{Nature}, 1990,
  \textbf{347}, 44\relax
\mciteBstWouldAddEndPuncttrue
\mciteSetBstMidEndSepPunct{\mcitedefaultmidpunct}
{\mcitedefaultendpunct}{\mcitedefaultseppunct}\relax
\EndOfBibitem
\bibitem[Wiggins and et~al.(2006)]{wiggins}
P.~A. Wiggins and et~al., \emph{Nature-Nanotecnology}, 2006, \textbf{1},
  137--41\relax
\mciteBstWouldAddEndPuncttrue
\mciteSetBstMidEndSepPunct{\mcitedefaultmidpunct}
{\mcitedefaultendpunct}{\mcitedefaultseppunct}\relax
\EndOfBibitem
\end{mcitethebibliography}
\bibliographystyle{rsc} %the RSC's .bst file

\end{document}